\documentclass[conference]{IEEEtran}
\IEEEoverridecommandlockouts
\usepackage{cite}
\usepackage{amsmath,amssymb,amsfonts}
\usepackage{algorithmic}
\usepackage{subcaption, graphicx}
\usepackage{textcomp}
\usepackage{xcolor}
\usepackage{booktabs, tabularx, makecell, array}
\usepackage{graphicx, subcaption, xcolor, tcolorbox} 
\usepackage{tikz}
\usepackage{url}
\usepackage{comment}

\newcommand{\hide}[1]{}
\def\BibTeX{{\rm B\kern-.05em{\sc i\kern-.025em b}\kern-.08em
    T\kern-.1667em\lower.7ex\hbox{E}\kern-.125emX}}
\begin{document}

\title{On-Device Qwen2.5: Efficient LLM Inference with Model Compression and Hardware Acceleration}


\author{
    Maoyang Xiang$^{*}$, Ramesh Fernando$^{*}$, Bo Wang \\
    Singapore University of Technology and Design \\
    Email:  maoyang\_xiang@sutd.edu.sg, ramesh\_fernando@mymail.sutd.edu.sg, bo\_wang@sutd.edu.sg
\thanks{$^{*}$Equal contribution.}
}

\maketitle

\begin{abstract}
Transformer-based Large Language Models (LLMs) have significantly advanced AI capabilities but pose considerable challenges for deployment on edge devices due to high computational demands, memory bandwidth constraints, and energy consumption. This paper addresses these challenges by presenting an efficient framework for deploying the Qwen2.5-0.5B model on the Xilinx Kria KV260 edge platform, a heterogeneous system integrating an ARM Cortex-A53 CPU with reconfigurable FPGA logic. Leveraging Activation-aware Weight Quantization (AWQ) with FPGA-accelerated execution pipelines, the proposed approach enhances both model compression rate and system throughput. Additionally, we propose a hybrid execution strategy that intelligently offloads compute-intensive operations to the FPGA while utilizing the CPU for lighter tasks, effectively balancing the computational workload and maximizing overall performance. 
Our framework achieves a model compression rate of 55.08\% compared to the original model and produces output at a rate of 5.1 tokens per second, outperforming the baseline performance of 2.8 tokens per second. 
\end{abstract}

\begin{IEEEkeywords}
LLMs, edge AI, FPGA, acceleration
\end{IEEEkeywords}

\section{Introduction}\label{sec:intro}
Recent advancements in Large Language Models (LLMs) have spurred numerous opportunities across different domains \cite{bommasani2022opportunitiesrisksfoundationmodels}, attracting users from sectors such as healthcare~\cite{Peng_2023_medical}, robotics~\cite{zeng2023largelanguagemodelsrobotics}, biomedical~\cite{BioMedicalDataAnalysis2025}, music~\cite{yuan-etal-2024-chatmusician}, etc. However, the growing adoption of these models has imposed immense computational demands, particularly for state-of-the-art models like GPT-4~\cite{openai2024gpt4technicalreport}, DeepSeek-V3~\cite{deepseekai2024deepseekv3technicalreport}, and PaLM 2~\cite{anil2023palm2technicalreport}, which are typically confined to power-intensive data centers. In contrast, there is a rising need to deploy lightweight LLMs on edge devices to enable real-time responses in wireless-denied environments and enhance privacy protection~\cite{huaEdgeComputingWithAI_2023}. In this work, we propose an efficient framework to accelerate on-device Qwen2.5-0.5B model inference.

\begin{figure}[t]
\centering
\includegraphics[width=\linewidth]{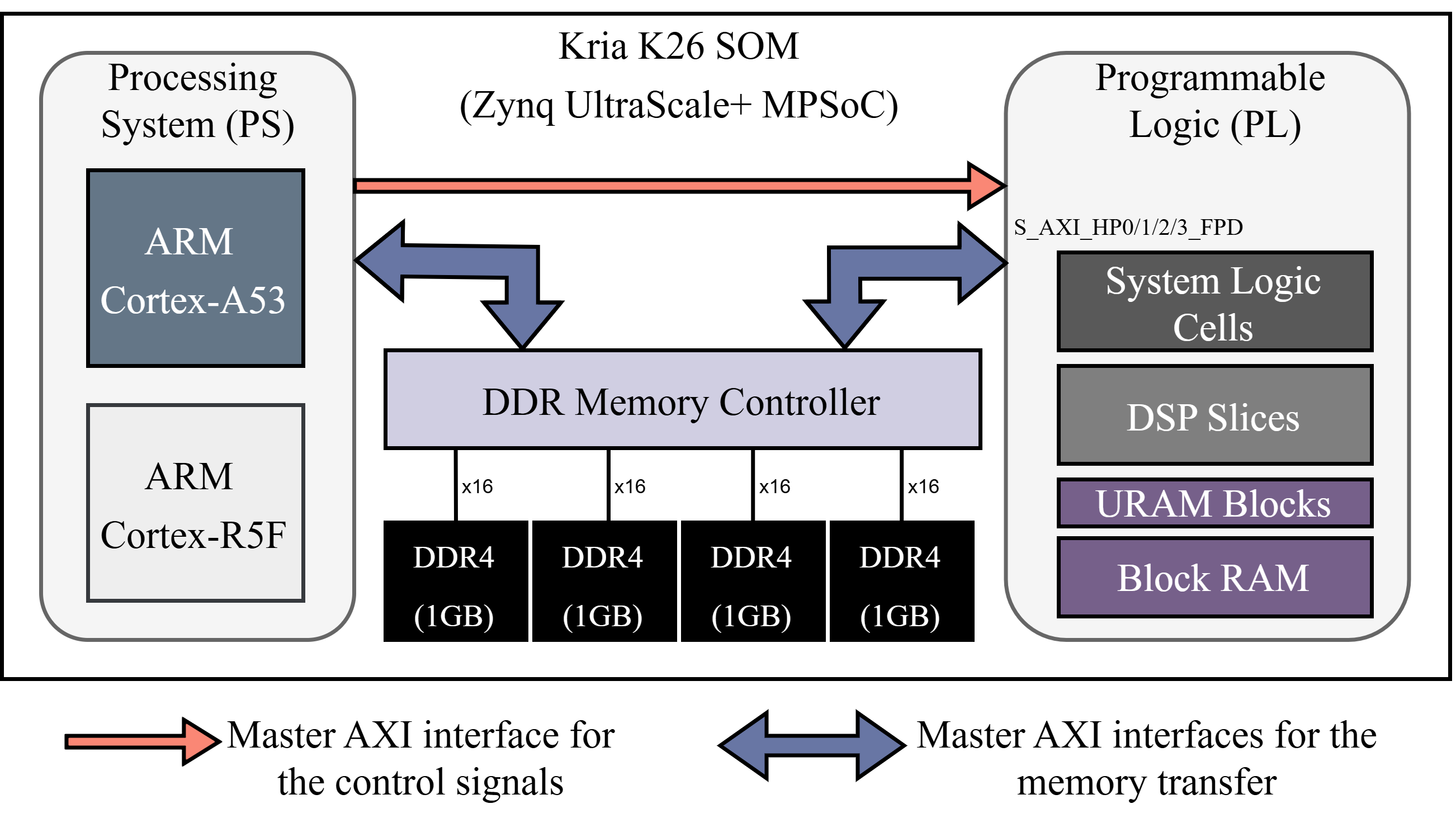}
\caption{Simplified hardware architecture of Xilinx Kria K26 system-on-module (SOM) adapted from~\cite{kria-k26-datasheet}.} 
\label{kria_k26_som}
\end{figure}

To implement it, we identify key challenges in deploying the model on resource-constrained edge devices and then employ the Xilinx Kria KV260 board, which incorporates the Kria K26 System-on-Module (SOM) (Fig.~\ref{kria_k26_som}) as the target platform. 
Specifically, we identify significant challenges deteriorating the efficiency of model inference on Kria KV260 through a detailed analysis of the Qwen2.5-0.5B model, including both linear and non-linear operations. 
\begin{itemize}
    \item Challenge 1: The limited memory capacity and bandwidth in the Programmable Logic (PL) side (e.g., Block RAMs, URAMs) restrict both the amount of model parameters that can be loaded and the efficiency of model loading, significantly impeding the deployment of on-device LLM inference.
    \item Challenge 2:  Matrix multiplications performed through Multiply-and-Accumulate (MAC) operations dominate the computational workload during inference, making them a performance bottleneck for LLM inference on edge devices.
\end{itemize}


\begin{figure*}[h]
    \centering
    \begin{tikzpicture}
    \node[anchor=south west, inner sep=0] (image) at (0,0)
    {\includegraphics[width=0.8\linewidth]{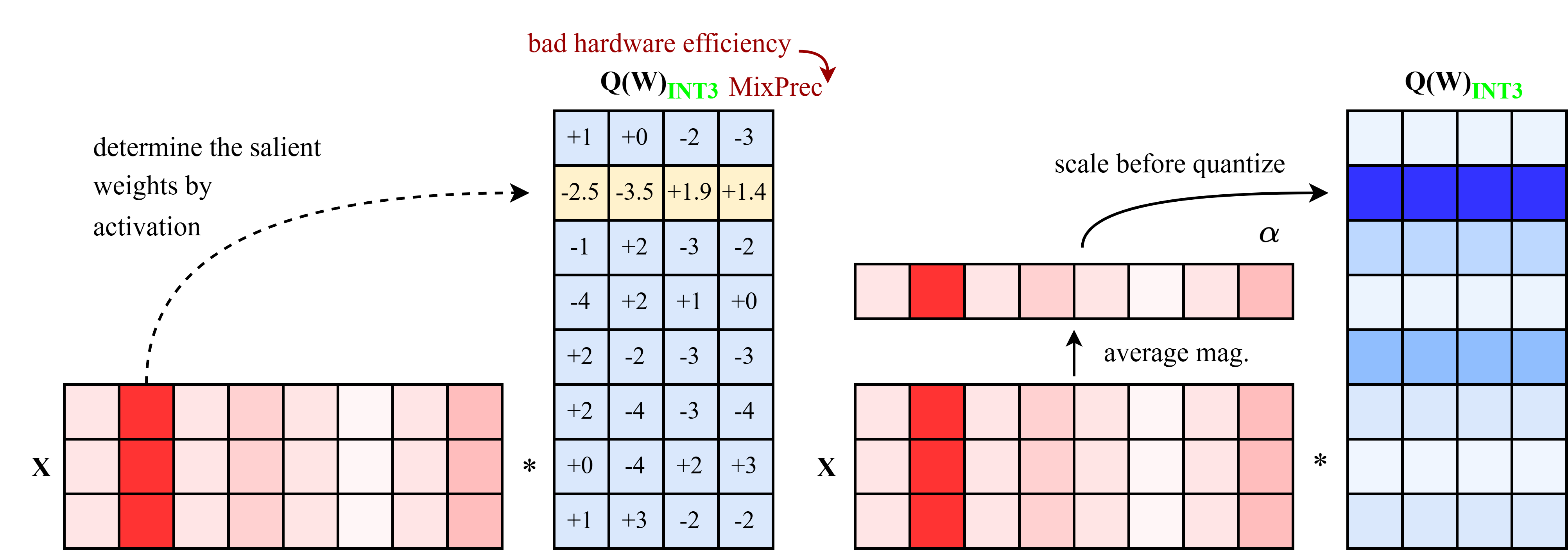}};
        \node at (4, -0.3) {(a) Keeping 1\% salient weights in FP16};
        \node at (11.5, -0.3) {(b) Scale the weights before quantization};
    \end{tikzpicture}
    \caption{AWQ uncovers that (a) keeping only 1\% of salient weights in FP16 can achieve a similar accuracy compared to the original model but not hardware friendly; (b) following activation awareness and performing per-channel scaling can protect the salient weights and reduce quantization error. The figure is adapted from work~\cite{lin2023awq}.}
    \label{fig-awq-methodology}
\end{figure*}

In this work, we address the challenges and achieve an excellent balance between accuracy, compression ratio, and tokens per second. The key contributions of our work are as follows.
\begin{itemize}
    \item We exploit a model compression technique at the software level to optimize memory bandwidth utilization and minimize data transfer between off-chip memory and processing elements.
    \item We leverage the VFP4 floating point unit in ARM Cortex-A53 and parallelism of the FPGA fabric and successfully realize end-to-end inference of the Qwen2.5-0.5B model.
    \item We further design an accelerator that can expedite matrix multiplications in the workload with piplined weight unpacking and dequantization.
    \item We achieve a model compression rate of 55.1\% and obtain an output rate of 5.1 tokens per second, surpassing the baseline performance of 2.8 tokens per second.   
\end{itemize}

\section{Background and Related Work}\label{top-background-related-works}
This section provides an overview of the Qwen2.5 model architecture and the AWQ-based model compression technique.
\subsection{The Qwen2.5 model and the Transformer Architecture}\label{qwen-architecture}
The Transformer architecture~\cite{NIPS2017_3f5ee243} has become the foundation of modern large language models, primarily due to its encoder-decoder structure. In particular, the Qwen2.5 model series~\cite{qwen2, qwen2.5} adopts a decoder-only architecture that utilizes an auto-regressive approach for sequential text generation~\cite{touvron2023llama2openfoundation}. However, this inherently sequential decoding process poses significant computational challenges.

Specifically, the decoder architecture in the Qwen2.5 model consists of a prefill stage and a decode stage. The prefill stage is characterized by highly parallel matrix-matrix operations while the decode stage is dominated by sequential matrix-vector multiplications. Consequently, accelerating matrix multiplications becomes crucial for optimizing compute efficiency and minimizing latency, particularly for applications requiring real-time inference capabilities.  

\subsection{Model compression with AWQ}
Activation-aware Weight Quantization (AWQ)~\cite{lin2023awq} is a powerful model compression technique to significantly reduce memory footprint and lower memory bandwidth requirements. As illustrated in Fig.~\ref{fig-awq-methodology}, only 1\% of the weights can have a significant impact on model performance. Therefore, carefully preserving those salient weights can maintain performance and minimize quantization error. Specifically, AWQ performs per-channel scaling based on activation distribution, so that the entire weight matrix can be quantized to lower precisions (e.g., INT4 and INT3) but still enjoy high performance. AWQ does not rely on model training or reconstruction. It utilizes a grouping mechanism to share the same scaling parameters with a default Group Size (GS) of 128. This grouping allows AWQ to accurately capture variations in weight distributions across different channels, effectively reducing quantization error and maintaining high model accuracy despite using low-precision representations.
We believe such model compression is particularly advantageous in edge deployment scenarios, such as the KV260 platform which is constrained by a memory bandwidth of 19.2 GB/s. Our implementation extends AWQ's benefits through a customized weight packing scheme, ensuring efficient delivery of the quantized weights (i.e., qweights), scaling factors, and zero values to the Processing Elements. This packing scheme facilitates on-the-fly dequantization, optimizing both performance and compute efficiency. 

\subsection{Related Work}\label{sub-related-work}
The utilization of FPGA to accelerate LLM inference has garnered significant interest from both academia and industry due to their exceptional energy efficiency, reconfigurability, and flexibility. A recent study~\cite{Chen2024FPGA} has introduced specialized spatial accelerators designed for efficient FPGA-based LLM inference, revealing the potential of FPGA platforms to provide customized, high-performance solutions. Particularly, the work in~\cite{LlamaFHanXuFPGA} has achieved substantial improvements in both performance and power efficiency after deploying the TinyLlama 1.1B model~\cite{zhang2024tinyllamaopensourcesmalllanguage} on the Xilinx ZCU102 platform, demonstrating the feasibility of FPGA-based deployments in edge inference applications. 
\section{Proposed Software-Hardware Co-Optimization}\label{sec-proposed-cooptimization}
We perform an in-depth analysis of the computational load and delay associated with the Qwen2.5-0.5B model. 
Table~\ref{tab:booktabs} presents a latency breakdown of the Qwen2.5-0.5B model during inference when deployed on the PS of the KV260 platform, using the compiler optimizations to utilize all four ARM Cortex-A53 cores. It reveals that 91.6\% of the inference time is dominated by MAC, which are the fundamental operations in matrix-matrix and matrix-vector multiplications. Note that the reported time in Table~\ref{tab:booktabs} accounts for both computation and memory access latency. This observation implies two key opportunities to improve the inference efficiency, which are (1) expediting compute-intensive operations using an accelerator and (2) compressing weight parameters in a data block so that it can be optimally stored in memory and transmitted to the accelerator with minimal bandwidth, improving memory bandwidth utilization.

\subsection{Software Optimization}\label{sec-software-architecture}
\begin{figure}[h]
\centering
\includegraphics[width=\linewidth]{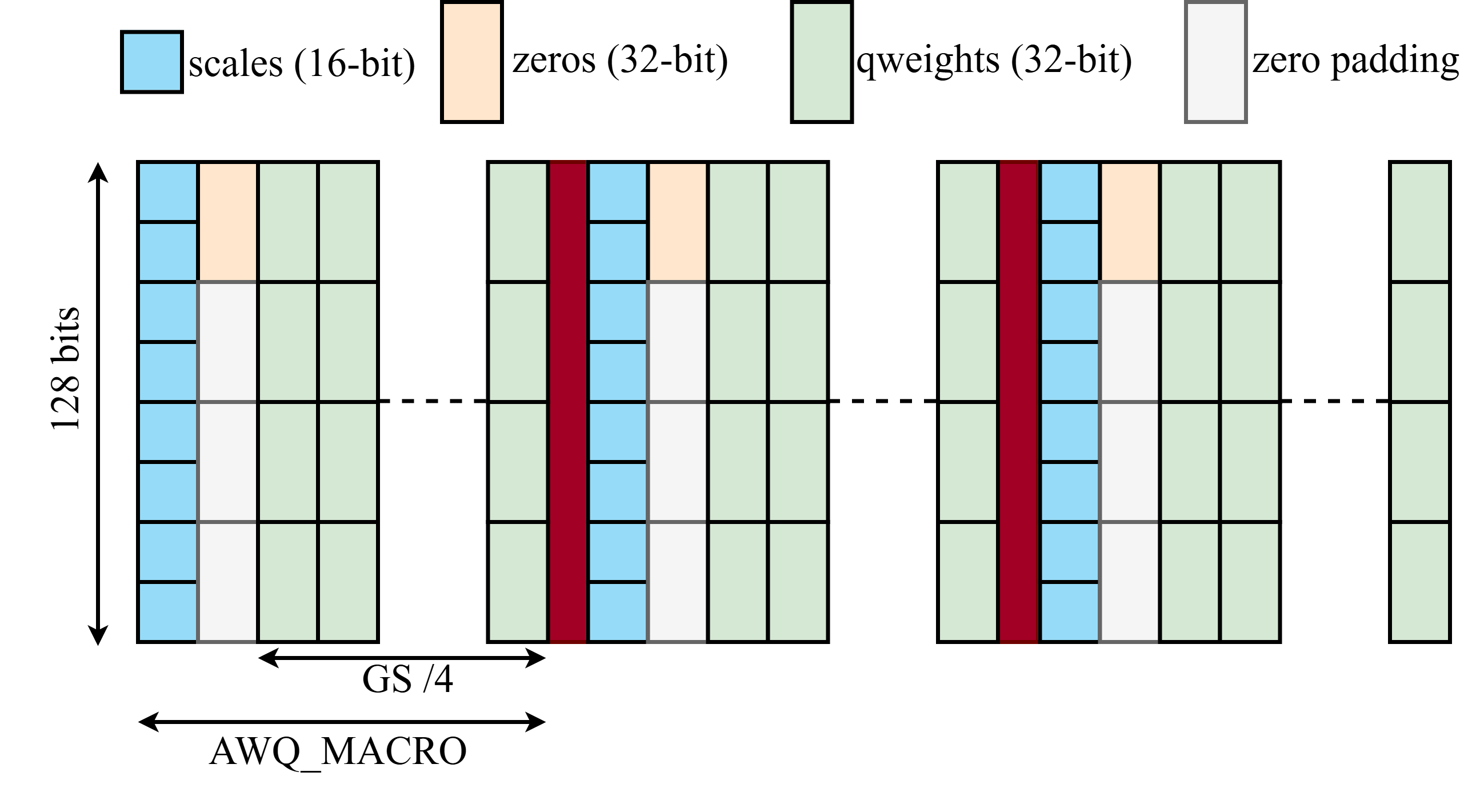}
\caption{Customized weight compression in the memory via AWQ\_MACRO, a block with scales, zeros, and quantized weights in INT4.} 
\label{weight_packing_memory}
\end{figure}

We exploit the AWQ framework~\cite{lin2023awq} to improve memory bandwidth utilization during inference. As Fig.~\ref{weight_packing_memory} illustrates, we pack the quantized weights, scales, and zeros in a data block. This enhances utilization of memory bandwidth when streaming data to the PL using 4 channels, each with 128-bit data. The accelerator in the PL can unpack weights, dequantize weights, and perform MAC operations using a pipelined architecture to exploit parallelism. We further modify the original weight mapping structure of AWQ to accommodate AWQ\_MACRO. As Fig.~\ref{weight_packing_memory} depicts, an AWQ\_MACRO packs $\text{Group Size (GS)} \times 8$ of quantized weights, 8 scales, and 8 zeros that are used to dequantize the weights in INT4 to FP32 prior to the pre-fill stage and decode stage. Since we need only 8 zero values in INT4 precision (i.e., 32 bits) to dequantize an AWQ\_MACRO, the remaining 96 bits out of the 128-bit strip in each macro are padded with zero.
\begin{table}[t]
    \centering
    \caption{Inference latency breakdown for Qwen2.5-0.5B model when deployed on the Processing System of the KV260 platform with -Ofast and SIMD optimizations.}
    \renewcommand{\arraystretch}{1.2} 
    \begin{tabularx}{\columnwidth}{X r r}
        \toprule
        Description & Time ($\mu s$) & Percentage (\%) \\ 
        \midrule
        \textbf{Linear Operations} & & \\
        Token Embedding copy + Layer init. & 32 & 0.20 \\ 
        Q/K/V Projection MAC operations & 1815 & \textbf{11.38} \\ 
        Q/K/V Bias addition & 35 & 0.22 \\
        Output projection + Residual Add & 839 & 5.26 \\
        MHA computation (Concatenation) & 42 & 0.26 \\
        FFN Gate Projection + Up Projection & 8148 & \textbf{51.08} \\
        FFN Down Projection + Residual Add & 4651 & \textbf{29.15} \\
        \midrule
        Overall MAC operations & 14,650 & \textbf{91.61} \\
        \midrule
        \textbf{Non-Linear Operations} & & \\
        Rotary Positional Encoding (RoPE) for Q/K & 87 & 0.54 \\
        Root Mean Square Normalization (RMSNorm) & 25 & 0.16 \\
        SiLU Activation + Element-wise Multiplication & 278 & 1.74 \\
        \midrule
        Total & 15,952 & 100 \\
        \bottomrule
    \end{tabularx}
    \label{tab:booktabs}
\end{table}

Our model compression approach incorporates several key features. First, data in an AWQ\_MACRO is streamed to the PL where dequantization and MAC operations are executed in a pipelined manner for improved efficiency. Second, the packing scheme ensures that the quantized weights are stored with their respective zero values and scale factors, enabling efficient per-channel dequantization. Lastly, the packing process is performed with a GS of 64 due to the higher accuracy score we achieved with the WNLI benchmark other than a GS of 128. The on-device inference of the Qwen2.5 model is implemented in C ~\cite{karpathy2023llama2c}. The model parameters (e.g., weights) are saved in a binary file while the model architecture is saved to a JSON file. Our approach is fully automated, allowing for seamlessly inference deployment on the KV260 platform with the AutoAWQ library~\cite{lin2023awq}, the binary file, and the JSON file.

    

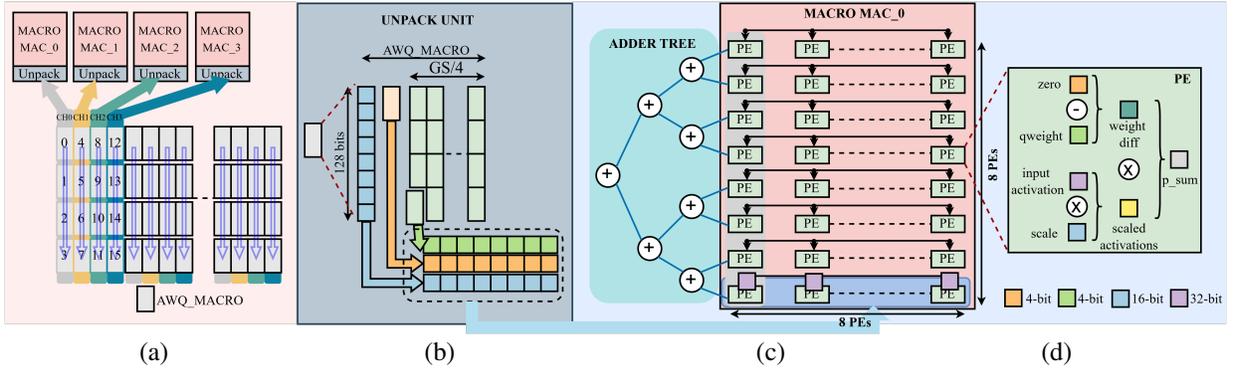
\begin{figure*}[h]
    \centering
    \begin{tikzpicture}
    \node[anchor=south west, inner sep=0] (image) at (0,0)
    {\includegraphics[width=0.9\linewidth]{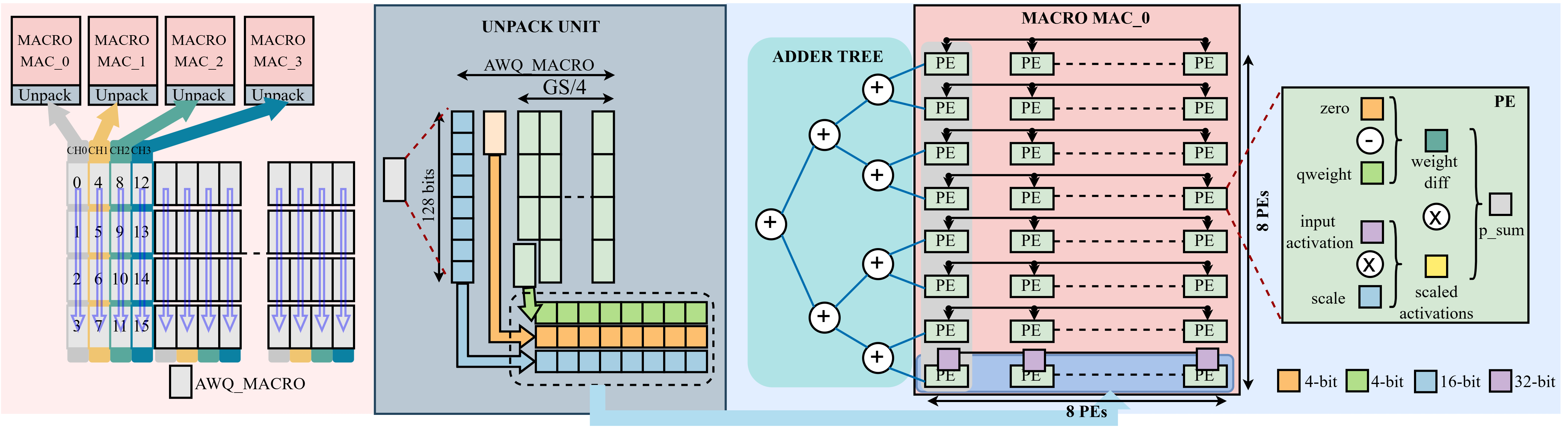}};
        \node at (2, -0.2) {(a)};
        \node at (5.8, -0.2) {(b)};
        \node at (10.2, -0.2) {(c)};
        \node at (14, -0.2) {(d)};
    \end{tikzpicture}
    \caption{(a) Order of loading AWQ\_MACRO to the MAC units using 4 AXI channels (b) Unpacking unit which unpacks the AWQ\_MACRO (c) MACRO MAC Unit with PE arrays, a sliding window in grey is used for accumulation of \textit{p\_sum} along the columns using the adder tree (d) Operations performed inside of a PE element.}
    \label{unpacking_dequantizing_mac}
\end{figure*}

\subsection{Hardware Optimization}\label{sec-hardware-architecture}
We optimize the hardware to support AWQ implementation as well as accelerate matrix computations. To efficiently handle data transfer, we utilize four Advanced eXtensible Interface (AXI) channels, streaming data directly into custom-designed MACRO\_MAC units. Fig.~\ref{unpacking_dequantizing_mac}(a) illustrates the memory access patterns, which ensure high throughput and structured data flow in the AXI channels. AWQ\_MACROs are carefully arranged in the memory as described in Section~\ref{sec-software-architecture}, which ensures a one-to-one correspondence between the dequantized values and their respective positions in the original weight matrix. The 4 MACRO\_MAC units can process data in parallel since the 4-AXI channels are independent.

The unpacking unit (Fig.~\ref{unpacking_dequantizing_mac}(b)) unpacks scales, zeros, and quantized weights (i.e., qweights) into a 128-bit format, respectively. The qweights block, which has been packed in a 32-bit integer format, is subsequently decomposed into 8 separate chunks in INT4 through a shift operation with a bit mask. The zero blocks undergo a similar disassembly process. It is worthwhile noting that the scales remain the original FP16 format throughout this stage. The unpacked qweights, zeros, and scales are then sent to the MACRO\_MAC units for weight dequantization before multiplying with input activations. We design an $8\times8$ Processing Elements (PE) array in PL so that the multiplications can be done in parallel. Fig.~\ref{unpacking_dequantizing_mac}(d) illustrates the operations in a PE, where zero value is subtracted from the qweight and the input activation is multiplied with the scale value. The partial sums from the PE array are accumulated subsequently with an adder tree. Once accumulations for an output channel are completed, the total sum will be transferred to the Processing System (PS) for non-linear operation computation.
\begin{table}[!h]
    \centering
    \caption{Synthesized resource utilization for the accelerator}
    \renewcommand{\arraystretch}{1.2} 
    \begin{tabularx}{\columnwidth}{|X|c|c|c|}
        \toprule
          & DSP & FF & LUT \\ 
        \midrule
        \makecell{4-MAC\_MACRO\\accelerator} & 384 (30\%) & 110,405 (47\%) & 96,553 (82\%) \\
        \bottomrule
    \end{tabularx}
    \label{tab:resource-consumption}
\end{table}

Table~\ref{tab:resource-consumption} presents the synthesis results of our accelerator, which operates at a frequency of 200 MHz. As the KV260 platform does not inherently support lower precision floating point formats, all MAC operations are performed in FP32.
\section{Evaluation}\label{sec-evaluation}
The performance of our software-hardware co-optimization framework has been evaluated against a baseline model in terms of accuracy, model size, throughput, and a benchmark score. The results are summarized in Table~\ref{tab:results}. Our compression approach results in a substantial reduction in model size. Specifically, the size decreases from 988 MB in the baseline model to 443.81 MB, achieving a reduction of 55.1\% in memory footprint. Moreover, the inference performance improves significantly from 2.8 tokens per second in the baseline to 5.1 tokens per second, nearly doubling the throughput. 

\newcolumntype{Y}{>{\centering\arraybackslash}X} 
\newcolumntype{Z}{>{\centering\arraybackslash}p{1.0\dimexpr\columnwidth/4\relax}} 
\renewcommand\cellalign{tl} 
\begin{table}[!h]
    \centering
    \caption{Evaluation Results} 
    \renewcommand{\arraystretch}{1.2} 
    \begin{tabularx}{\columnwidth}{|Z Y Y Y Y|}
        \toprule
        & \makecell{Accuracy\\(\centering{\%})} &\makecell{Model Size\\ (MB)} & \makecell{Tokens/s} & \makecell{Total \\ Score}\\ 
        \midrule
        Baseline & 64.79 & 988 & 2.80 & 0.4 \\ 
        Ours\\(AWQ GS=64) & 61.97 & 443.81 & 5.10$^*$ & 0.55 \\       
        \bottomrule
    \end{tabularx}
    \label{tab:results}
\end{table}
$^*$Based on co-simulation results.

\begin{equation}
\begin{aligned}
Tot_{1st} &= 0.4 \times \frac{Ratio_{accuracy}}{\mathrm{MAX}(Ratio_{accuracy})} \\
          &\quad + 0.2 \times \frac{Ratio_{memory}}{\mathrm{MAX}(Ratio_{memory})} \\
          &\quad + 0.2 \times \frac{Ratio_{throughput\_P}}{\mathrm{MAX}(Ratio_{throughput\_P})} \\
          &\quad + 0.2 \times \frac{Ratio_{throughput\_D}}{\mathrm{MAX}(Ratio_{throughput\_D})}
\end{aligned}
\label{eqn:total-score}
\end{equation}
Moreover, we adopt Equation~\eqref{eqn:total-score} to generate the benchmark score for a comprehensive evaluation. Specifically, the accuracy ratio represents how accurately the model can perform with the WNLI~\cite{elfwing2017sigmoidweightedlinearunitsneural} benchmark. The memory ratio shows how well the model can be compressed compared to the original model size. The throughput at the prefill stage and the decode stage represents how efficiently the tokens can be processed at each stage, respectively. We achieve a score of 0.55 compared to the baseline score of 0.4. 

\section{Conclusion}
Our work presents an end-to-end inference framework that harnesses the synergetic strength of the Processing System and Programmable Logic on an FPGA platform to enable efficient deployment of the Qwen2.5-0.5B model. Our framework achieves a high compression rate that reduces memory footprint and facilitates higher memory bandwidth utilization when transferring the model parameters from off-chip memory to the PS. Moreover, we propose a hardware accelerator by utilizing a PE array with adder tree accumulation, expediting the performance of matrix multiplications.
The gains in both model size and throughput make it highly beneficial for LLM inference deployment in resource-constrained environments.


\begin{thebibliography}{10}
\bibitem{bommasani2022opportunitiesrisksfoundationmodels}
R.~Bommasani, D.~A. Hudson, E.~Adeli, R.~Altman, S.~Arora, S.~von Arx, {\em et~al.}, ``On the opportunities and risks of foundation models,'' 2022.
\newblock Available at \url{https://arxiv.org/abs/2108.07258}.

\bibitem{Peng_2023_medical}
C.~Peng, X.~Yang, A.~Chen, K.~E. Smith, N.~PourNejatian, A.~B. Costa, C.~Martin, M.~G. Flores, Y.~Zhang, T.~Magoc, G.~Lipori, D.~A. Mitchell, N.~S. Ospina, M.~M. Ahmed, W.~R. Hogan, E.~A. Shenkman, Y.~Guo, J.~Bian, and Y.~Wu, ``A study of generative large language model for medical research and healthcare,'' {\em npj Digital Medicine}, vol.~6, Nov. 2023.

\bibitem{zeng2023largelanguagemodelsrobotics}
F.~Zeng, W.~Gan, Y.~Wang, N.~Liu, and P.~S. Yu, ``Large language models for robotics: A survey,'' 2023.
\newblock Available at \url{https://arxiv.org/abs/2311.07226}.

\bibitem{BioMedicalDataAnalysis2025}
W.~Lan, Z.~Tang, M.~Liu, Q.~Chen, W.~Peng, Y.~P. Chen, and Y.~Pan, ``The large language models on biomedical data analysis: A survey,'' {\em IEEE Journal of Biomedical and Health Informatics}, pp.~1--13, 2025.

\bibitem{yuan-etal-2024-chatmusician}
R.~Yuan, H.~Lin, Y.~Wang, Z.~Tian, S.~Wu, T.~Shen, {\em et~al.}, ``{C}hat{M}usician: Understanding and generating music intrinsically with {LLM},'' in {\em Findings of the Association for Computational Linguistics: ACL 2024}, (Bangkok, Thailand), pp.~6252--6271, Association for Computational Linguistics, Aug. 2024.

\bibitem{openai2024gpt4technicalreport}
OpenAI, A.~Josh, A.~Steven, A.~Sandhini, A.~Lama, A.~Ilge, A.~Florencia, A.~Diogo, A.~Janko, A.~Sam, A.~Shyamal, A.~Red, B.~Igor, B.~Suchir, B.~Valerie, B.~Paul, B.~Haiming, B.~Mohammad, B.~Jeff, and Z.~Barret, ``Gpt-4 technical report,'' 03 2023.
\newblock Available at \url{10.48550/arXiv.2303.08774}.

\bibitem{deepseekai2024deepseekv3technicalreport}
DeepSeek-AI, ``Deepseek-v3 technical report,'' 2024.
\newblock Available at \url{https://arxiv.org/abs/2412.19437}.

\bibitem{anil2023palm2technicalreport}
R.~Anil, A.~M. Dai, O.~Firat, M.~Johnson, D.~Lepikhin, A.~Passos, {\em et~al.}, ``Palm 2 technical report,'' 2023.
\newblock Available at \url{https://arxiv.org/abs/2305.10403}.

\bibitem{huaEdgeComputingWithAI_2023}
H.~Hua, Y.~Li, T.~Wang, N.~Dong, W.~Li, and J.~Cao, ``Edge computing with artificial intelligence: A machine learning perspective,'' {\em ACM Comput. Surv.}, vol.~55, Jan. 2023.

\bibitem{kria-k26-datasheet}
{AMD Xilinx}, {\em Kria K26 SOM Data Sheet}.
\newblock AMD, 2021.
\newblock DS987 (v1.5), May 18, 2021.

\bibitem{lin2023awq}
J.~Lin, J.~Tang, H.~Tang, S.~Yang, W.-M. Chen, W.-C. Wang, G.~Xiao, X.~Dang, C.~Gan, and S.~Han, ``Awq: Activation-aware weight quantization for llm compression and acceleration,'' in {\em MLSys}, 2024.

\bibitem{NIPS2017_3f5ee243}
A.~Vaswani, N.~Shazeer, N.~Parmar, J.~Uszkoreit, L.~Jones, A.~N. Gomez, L.~u. Kaiser, and I.~Polosukhin, ``Attention is all you need,'' in {\em Advances in Neural Information Processing Systems} (I.~Guyon, U.~V. Luxburg, S.~Bengio, H.~Wallach, R.~Fergus, S.~Vishwanathan, and R.~Garnett, eds.), vol.~30, Curran Associates, Inc., 2017.

\bibitem{qwen2}
A.~Yang, B.~Yang, B.~Hui, B.~Zheng, B.~Yu, C.~Zhou, C.~Li, C.~Li, D.~Liu, F.~Huang, {\em et~al.}, ``Qwen2 technical report,'' {\em arXiv preprint arXiv:2407.10671}, 2024.

\bibitem{qwen2.5}
Q.~Team, ``Qwen2.5: A party of foundation models,'' September 2024.
\newblock Available at \url{https://qwenlm.github.io/blog/qwen2.5/}.

\bibitem{touvron2023llama2openfoundation}
H.~Touvron, L.~Martin, K.~Stone, P.~Albert, A.~Almahairi, Y.~Babaei, {\em et~al.}, ``Llama 2: Open foundation and fine-tuned chat models,'' 2023.
\newblock Available at \url{https://arxiv.org/abs/2307.09288}.

\bibitem{Chen2024FPGA}
H.~Chen, J.~Zhang, Y.~Du, S.~Xiang, Z.~Yue, N.~Zhang, Y.~Cai, and Z.~Zhang, ``Understanding the potential of {FPGA}-based spatial acceleration for large language model inference,'' {\em ACM Transactions on Reconfigurable Technology and Systems}, vol.~18, no.~1, pp.~1--23, 2024.

\bibitem{LlamaFHanXuFPGA}
H.~Xu, Y.~Li, and S.~Ji, ``Llamaf: An efficient llama2 architecture accelerator on embedded fpgas,'' in {\em 2024 IEEE 10th World Forum on Internet of Things (WF-IoT)}, pp.~1--7, 2024.

\bibitem{zhang2024tinyllamaopensourcesmalllanguage}
P.~Zhang, G.~Zeng, T.~Wang, and W.~Lu, ``Tinyllama: An open-source small language model,'' 2024.
\newblock Available at \url{https://arxiv.org/abs/2401.02385}.

\bibitem{karpathy2023llama2c}
A.~Karpathy, ``llama2.c: Inference of llama2 in one file of pure c,'' 2023.
\newblock Available at \url{https://github.com/karpathy/llama2.c}.

\bibitem{elfwing2017sigmoidweightedlinearunitsneural}
S.~Elfwing, E.~Uchibe, and K.~Doya, ``Sigmoid-weighted linear units for neural network function approximation in reinforcement learning,'' 2017.
\newblock Available at \url{https://arxiv.org/abs/1702.03118}.
\end{thebibliography}

\end{document}